\documentclass{elsart}

\usepackage{graphicx,amsmath,amsfonts,amssymb}

\newcommand{\ket}[1]{\ensuremath{|{#1\rangle}}} 
\newcommand{\bra}[1]{\ensuremath{{\langle #1}|}}

\newcommand{\ketbra}[2]{\ensuremath{|{#1 \rangle}{\langle #2}|}}
\newcommand{\avg}[1]{\ensuremath{{\langle#1\rangle}}} 
\newcommand{\E}{\ensuremath{\text{e}}}

\newcommand{\I}{\ensuremath{\text{i}}}
\providecommand{\abs}[1]{\left\lvert#1\right\rvert}
\newcommand{\op}[1]{\hat{#1}}
\renewcommand{\eqref}[1]{Eq.~(\ref{#1})}

\begin{document}

\begin{frontmatter}



\title{Equivalence between Bell inequalities and quantum Minority game}


\author[ad1]{Adrian P. Flitney\corauthref{cor}}
\corauth[cor]{Corresponding author.}
\ead{aflitney@unimelb.edu.au}
\author[ad1]{Maximilian Schlosshauer}
\author[ad2]{Christian Schmid} 
\author[ad3]{Wieslaw Laskowski}
\author[ad1,ad4]{Lloyd C. L. Hollenberg}
\address[ad1]{School of Physics, University of Melbourne, VIC 3010, Australia}
\address[ad2]{Max-Planck-Institute for Quantum Optics, D-85748 Garching, Germany}
\address[ad3]{Institute of Theoretical Physics and Astrophysics,
  University of Gda\'{n}sk, PL-80-952 Gda\'{n}sk, Poland}
\address[ad4]{Centre of Excellence for Quantum Computer Technology, University of Melbourne, Australia}

\begin{abstract}
  We show that, for a continuous set of entangled four-partite states,
  the task of maximizing the payoff in the symmetric-strategy
  four-player quantum Minority game is equivalent to maximizing the
  violation of a four-particle Bell inequality with each observer
  choosing the same set of two dichotomic observables.  We conclude
  the existence of direct correspondences between (i) the payoff rule
  and Bell inequalities, and (ii) the strategy and the choice of
  measured observables in evaluating these Bell inequalities.
  We also show that such a correspondence between Bell polynomials (in a single plane) and four-player,
  symmetric, binary-choice quantum games is unique to the four-player quantum Minority game
  and its ``anti-Minority'' version.
  This indicates that the four-player Minority game not only plays a special role among quantum games but
  also in studies of Bell-type quantum nonlocality.
\end{abstract}

\begin{keyword}
Quantum game theory; Bell inequalities; Minority game

\PACS 03.67.-a, 02.50.Le
\end{keyword}
\end{frontmatter}


\maketitle

\section{Introduction}
The experimentally observed violations of Bell inequalities reflect
deep aspects of realism and locality in
nature~\cite{groblacher07,leggett03} supporting the quantum-mechanical
description of correlations between spatially separated systems.
Quantum game theory, on the other hand, is an active
branch of quantum information theory yet seems far removed
from such physical truths, recent implementations
notwithstanding~\cite{prevedel07,schmid07}.  In
this work we show for the first time a remarkable equivalence between aspects of
these seemingly disparate corners of quantum theory.
While one may expect a correspondence at some level between quantum game theory
and non-classical correlations through Bell inequalities~\cite{enk02},
the exact equivalence we have uncovered is rather surprising.

Using an entangled state resource that utilizes a superposition of the
GHZ state and products of EPR pairs, we demonstrate equivalence
between the optimal symmetric strategy payoffs for a quantum Minority
game (QMG) and the violation of the
Mermin--Ardehali--Belinskii--Klyshko (MABK)
inequality~\cite{mermin90,ardehali92,belinskii93} for the initial
state.  Although previous publications have drawn a comparison between
so-called non-local games and Bell
inequalities~\cite{silman07,pawlowski07}, non-local games are
cooperative tasks for teams of remote players and not games in the von
Neumann sense.  Our results are the first direct equivalence between
the payoffs in a competitive quantum game and violation of Bell
inequalities.

\begin{figure}[ht]
\setlength{\unitlength}{0.5pt}
\begin{center}
\begin{picture}(565,350)(0,0)

\put(0,0){
%
%
	\begin{picture}(215,300)(0,-60)
	\put(25,100){\oval(50,100)}
	\put(50,60){\vector(1,-1){30}}
	\put(80,30){\line(1,-1){30}}
	\put(50,85){\vector(2,-1){30}}
	\put(80,70){\line(2,-1){30}}
	\put(50,115){\vector(2,1){30}}
	\put(80,130){\line(2,1){30}}
	\put(50,140){\vector(1,1){30}}
	\put(80,170){\line(1,1){30}}
	\multiput(115,0)(0,55){2}{\circle{10}}
	\multiput(115,147)(0,55){2}{\circle{10}}
	\put(105,178){$\scriptstyle \op{M}_1$}
	\put(105,123){$\scriptstyle \op{M}_2$}
	\put(105,65){$\scriptstyle \op{M}_3$}
	\put(105,10){$\scriptstyle \op{M}_4$}

	\put(120,202){\vector(1,0){35}}
	\put(155,202){\line(1,0){25}}
	\put(120,147){\vector(1,0){35}}
	\put(155,147){\line(1,0){25}}
	\put(120,55){\vector(1,0){35}}
	\put(154,55){\line(1,0){25}}
	\put(120,0){\vector(1,0){35}}
	\put(155,0){\line(1,0){25}}

	\multiput(185,0)(0,55){2}{\circle{10}}
	\multiput(185,147)(0,55){2}{\circle{10}}

	\put(2,95){$\scriptstyle \ket{\psi_\text{in}}$}

	\put(118,100){\oval(30,230)}
	\put(185,100){\oval(30,230)}

	\put(60,225){\shortstack{\scriptsize player\\ \scriptsize strategies}}
	\put(110,-38){$\Downarrow$}
	\put(40,-60){\scriptsize payoff polynomial}
	\put(160,245){\scriptsize measurement}
	\put(160,225){$\scriptstyle \{\op{P}_0, \op{P}_1 \}$}

	\end{picture}}

\put(215,0){
%
%
	\begin{picture}(150,240)(0,20)
	\put(0,220){$\Longrightarrow$}
	\put(50,200){\framebox(90,50){\shortstack{\scriptsize optimal \\ \scriptsize  payoff}}}
	\put(85,170){$\Updownarrow$}
	\put(50,100){\framebox(90,50){\shortstack{\scriptsize maximal \\ \scriptsize  violation}}}
	\put(150,120){$\Longleftarrow$}
	\put(85,25){\vector(-1,0){70}}
	\put(85,25){\vector(1,0){70}}
	\end{picture}}

\put(365,0){
%
%
	\begin{picture}(200,300)(15,-60)
	\put(175,100){\oval(50,100)}
	\put(150,60){\vector(-1,-1){30}}
	\put(120,30){\line(-1,-1){30}}
	\put(150,85){\vector(-2,-1){30}}
	\put(120,70){\line(-2,-1){30}}
	\put(150,115){\vector(-2,1){30}}
	\put(120,130){\line(-2,1){30}}
	\put(150,140){\vector(-1,1){30}}
	\put(120,170){\line(-1,1){30}}
	\multiput(85,0)(0,55){2}{\circle{10}}
	\multiput(85,147)(0,55){2}{\circle{10}}
	\put(60,225){$\scriptstyle \{ \op{A}_1, \op{A}_2 \}$}

	\put(155,95){$\scriptstyle \ket{\psi_\text{in}}$}

	\put(85,100){\oval(30,230)}
	\put(80,-38){$\Downarrow$}
	\put(40,-60){\scriptsize Bell polynomial}
	\put(60,245){\scriptsize local observables}

	\end{picture}}

\end{picture}
\end{center}
\caption{\label{fig:qmg} Schematic showing the equivalences between
  (\emph{left}) a four-player quantum Minority game (QMG) and
  (\emph{right}) a four-partite Bell inequality.  In the QMG, each
  player acts on one qubit from an entanglement resource
  $\ket{\psi_\text{in}}$ with a local (unitary) strategy $\op{M}_i$,
  followed by a projective measurement in the computational basis.
  The payoff matrix and the chosen strategies determine a payoff
  polynomial.  In the Bell inequality, observers choose local
  observables $\{\op{A}_1, \op{A}_2 \}$ to make measurements on each
  qubit of the same entanglement resource.  Optimizing the payoff in
  the game is equivalent to maximizing the violation of the MABK-type
  Bell inequality.}
\end{figure}
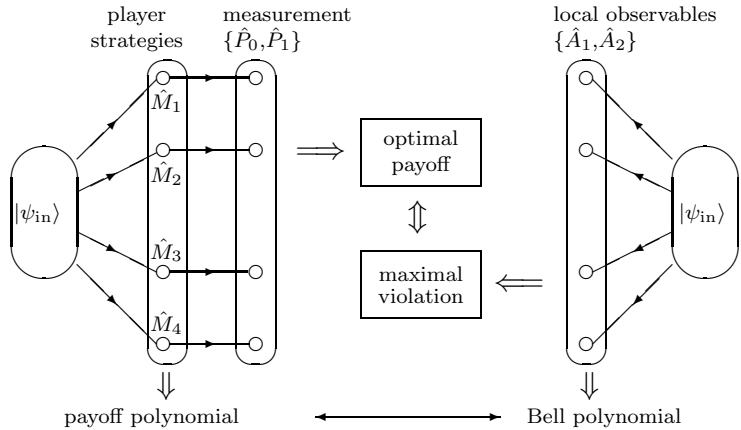

The classical Minority game~\cite{challet97} is a simple multi-player
game for studying coordination amongst a group of agents in the
absence of communication.
In each round, the agents
must independently select one of two options, labeled `0' and `1'.
Those that select the option chosen by the
minority win and are awarded a payoff of one unit,
while the others receive zero payoff.
In a one-off classical game the players can do no better than using the mixed strategy of
selecting each of the two options with equal probability.
When there are an even number of players this can result in no minority and hence
zero payoff to all players.

While game theory is the mathematical language of competitive (classical) interactions,
quantum game theory is the natural language to consider competitive situations
in a quantum information setting.
One exploits the classical framework as a base for finding new ways of understanding and using entanglement
in this context.
Although playing a game using entanglement as a resource is not
the same as playing the underlying classical game,
sharing entanglement is less strong than explicit cooperation.
In quantum versions of the Minority game~\cite{benjamin01b,chen04,flitney07a,flitney07b},
for even numbers of
players it has been shown~\cite{benjamin01b,chen04} that the
probability of getting no winners in the final state can be reduced,
to the benefit of all players.  For small numbers of players the QMG
is amenable to experimental implementation using multi-photon
entangled states~\cite{schmid07}.

Quantization of the Minority game proceeds as follows and is
shown schematically on the left-hand side of Fig.~\ref{fig:qmg}.  Each
of $N$ players receives one qubit from a known entangled state.  The
players' strategy is their choice of local unitary operator to apply
to their qubit,
\begin{equation}
\label{eq:su2}
\op{M}(\theta, \beta_1, \beta_2) = \left( \begin{array}{cc} e^{\I
      \beta_1} \cos(\theta/2) & \I e^{\I
      \beta_2} \sin(\theta/2) \\
    \I e^{-\I \beta_2} \sin(\theta/2) & e^{-\I \beta_1} \cos(\theta/2)
	\end{array} \right),
\end{equation}
where $\theta \in [0, \pi], \, \beta_1, \beta_2 \in [-\pi, \pi]$.  No
communication between the players is permitted and coherence is
maintained until after the players' actions.  Then the qubits are
measured in the computational basis and payoffs are awarded using the
usual payoff scheme. If the initial state is a GHZ state, the scheme
described is equivalent to the protocol of Eisert {\em et
  al.}~\cite{eisert99}, since the unentangling gate in Eisert's
protocol has no effect on the payoffs for the Minority
game~\cite{benjamin01b}.  The scheme is also consistent with the
generalized quantum game formalisms of Lee and Johnson~\cite{lee03b}
and Gutoski and Watrous~\cite{gutoski06}.

\section{Pareto optimal strategies and Bell inequalities}
In the current context we will be concerned with the Pareto-optimal
(PO) strategy profile, one from which no player can improve their
result without another being worse off.  We only consider the
situation where all players use the same strategy, since asymmetric
strategy profiles are problematic to achieve in the absence of
communication.  The symmetric PO result gives the maximal payoff that
is fair to all players.

Existing works have concentrated on using a GHZ state as the
entanglement resource for the game, however, we shall consider the
more general initial state
\begin{equation}
  \label{eq:initial}
  \ket{\psi_\text{in}} = \alpha \ket{\text{GHZ}} \:+\: \sqrt{1-\alpha^2} \,
  \ket{\text{EPR}}_{\text{AB}} \otimes \ket{\text{EPR}}_{\text{CD}},
\end{equation}
where $\ket{\text{GHZ}} \equiv \left(\ket{0000}+\ket{1111}\right) /
\sqrt{2}$, $\ket{\text{EPR}} \equiv \left(\ket{01} + \ket{10} \right)
/ \sqrt{2}$, and $\alpha \in [0, 1]$.  This state can be created
experimentally for arbitrary $\alpha$ using photons produced from down
conversion~\cite{schmid07,weinfurter01}.

Below we show that the payoff observable for the Minority game,
when transformed by the PO strategy profile,
gives rise to a Bell-like polynomial that,
when evaluated for the initial state \eqref{eq:initial},
is identical for all $\alpha$
(up to an arbitrary scaling factor)
to the maximal value of the MABK Bell polynomial.

\begin{figure}
\begin{center}
\includegraphics[scale=0.7]{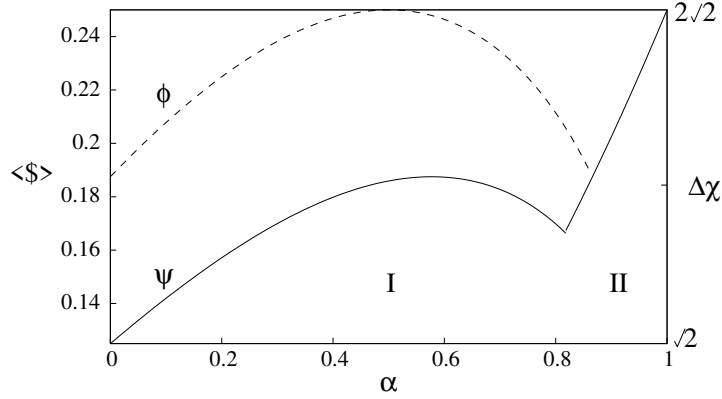}
\end{center}
\caption{\label{fig:payoffs} {\it Left scale:} Pareto optimal payoffs
  $\avg{\$_<}$ (region I) and $\avg{\$_>}$ (region II) in a four-player quantum
  Minority game with the initial states $\ket{\psi_\text{in}}$ [\eqref{eq:initial}]
  and $\ket{\phi_\text{in}}$ [\eqref{eq:initial2}].
  {\it Right scale:} Maximal violation of the MABK Bell equality
  \eqref{eq:mabk_in} for the same states.
  For measurement schemes, see text.
The {\it quantum fulcrum} between regions I and II
occurs at $\alpha = \sqrt{\frac{2}{3}}$ for $\ket{\psi_\text{in}}$
and $\alpha = \sqrt{\frac{3}{4}}$ for $\ket{\phi_\text{in}}$.}
\end{figure}

In the game, the state prior to measurement is
$\ket{\psi_\text{final}} = \op{M}_1 \otimes \op{M}_2 \otimes \op{M}_3
\otimes \op{M}_4 \, \ket{\psi_\text{in}}$, where the $\op{M}_i$ are
the strategies chosen by the players.  For symmetric strategy profiles
only the difference between the phases $\beta_1, \beta_2$ of
\eqref{eq:su2} is relevant.  We shall therefore, without loss of
generality, set $\beta \equiv \beta_1 = -\beta_2$.  A necessary and
sufficient condition for $\op{M}(\theta^{*}, \beta^{*}, -\beta^{*})$
to be a symmetric PO strategy is $\left\langle \$ \left(
    \op{M}(\theta^{*}, \beta^{*}, -\beta^{*})^{\otimes 4} \right)
\right\rangle \ge \left\langle \$ \left( \op{M}(\theta, \beta,
    -\beta)^{\otimes 4} \right) \right\rangle \; \forall \, \theta,
\beta$, where $ \$ $ represents the payoff to any one of the four
players for the indicated strategy profile.  When all players select
the strategy $\hat{M}(\theta, \beta, -\beta)$ for some $\theta, \,
\beta$ to be determined, the average payoff to each player is
$\avg{\$} = \frac{\sin^2 \theta}{32} [ 8 - 2 \alpha^2 + 8 \alpha
\sqrt{2 - 2 \alpha^2} \, \cos 4 \beta - 2 \alpha^2 \cos 8 \beta + 2(4
- 3 \alpha^2) \cos 2\theta + 8 \alpha \sqrt{2 - 2 \alpha^2} \, \cos 4
\beta \, \cos 2 \theta + 2 \alpha^2 \cos 8 \beta \, \cos 2 \theta
\,]$.  A local maximum or minimum in the value of the payoff will have
$d \langle \$ \rangle/d \theta = d \langle \$ \rangle/d \beta = 0$.
From these conditions we find the strategy that maximizes the payoff
to be
\begin{subequations}
\label{eq:dflksjg1}
\begin{align}
  \op{M}_> &= \frac{1}{\sqrt{2}} \left( \begin{array}{cc} \E^{\I
        \pi/8} & \I \E^{-\I \pi/8} \\ \I \E^{\I \pi/8} & \E^{-\I
        \pi/8} \end{array} \right) & \text{for $\alpha \ge
    \sqrt{\frac{2}{3}}$}, \\
  \op{M}_< &= 
  \left( \begin{array}{cc} \cos(\pi/8) & \I \sin(\pi/8) \\ \I
      \sin(\pi/8) & \cos(\pi/8) \end{array} \right) & \text{for
    $\alpha \le \sqrt{\frac{2}{3}}$},
\end{align}
\end{subequations}
with  expected payoffs $\avg{\$_>} = \frac{1}{4}\alpha^2$
and
\begin{equation}
 \avg{\$_<} = \underbrace{\frac{1}{16}}_{\op{I} \op{I} \op{I} \op{I}, \ldots}
	\:+\: \frac{1}{16} \Bigl( \underbrace{1-\alpha^2}_{-\op{Y} \op{Y} \op{Z} \op{Z}, \ldots}
	+\, 2 \: \underbrace{\sqrt{2} \alpha \sqrt{1-\alpha^2} }_{-\hat{Y} \hat{Z} \hat{Y} \hat{Z}, \ldots} \Bigr),
\end{equation}
where the contributions from the corresponding measurement
correlations $\bra{\psi_\text{in}} \cdots \ket{\psi_\text{in}}$ are
explicitly given under the braces [see \eqref{eq:payoff_ob1}].  The
payoffs as a function of $\alpha$ are shown in Fig.~\ref{fig:payoffs}.
At $\alpha = \sqrt{\frac{2}{3}}$, we find that, disregarding phases,
the states $\op{M}_>^{\otimes 4} \ket{\psi_\text{in}}$ and
$\op{M}_<^{\otimes 4} \ket{\psi_\text{in}}$ contain exactly the same
terms, all with identical prefactors.  Hence $\alpha =
\sqrt{\frac{2}{3}}$ is the boundary between the EPR- and the
GHZ-dominated regions.  This point is a {\em fulcrum} in the quantum
state where there is a switch in optimal strategy from $\op{M}_<$ to
$\op{M}_>$.

Let us write down a \emph{payoff observable} $\op{\$}$ as a sum of
projection operators on the winning states,
\begin{align}
\label{eq:payoff_op}
\op{\$} &= \frac{1}{4} \bigl( \op{P}_0^{(1)} \otimes \op{P}_1^{(2)}
\otimes \op{P}_1^{(3)} \otimes \op{P}_1^{(4)} +
\text{permutations} \notag \\
& \quad + \op{P}_0^{(1)} \otimes \op{P}_0^{(2)} \otimes \op{P}_0^{(3)}
\otimes \op{P}_1^{(4)} + \text{permutations} \bigr),
\end{align}
where $\op{P}_j^{(k)}$ is the projector $\ketbra{j}{j}, \; j=0,1$, for
the $k$th qubit.  The average payoff is then given by
\begin{align}
\label{eq:payoff_av}
  \langle \$ \rangle = \bra{\psi_\text{in}} \op{M}^{\dagger\otimes 4} \,
  \op{\$} \, \op{M}^{\otimes 4} \ket{\psi_\text{in}}.
\end{align}
Using $\op{M}_<^{\dagger} \, \op{P}_{0/1} \, \op{M}_< =
\frac{1}{2} \bigl(\op{I} \pm \op{B}_{zy} \bigr)$
with $\op{B}_{zy} \equiv \frac{1}{\sqrt{2}} (\op{Z} - \op{Y})$,
the payoff observable
\eqref{eq:payoff_op} transformed by the optimal strategy for
$\alpha < \sqrt{\frac{2}{3}}$ is
\begin{align}
\label{eq:payoff_ob1}
\op{\$}_< &\equiv \op{M}_<^{\dagger\otimes 4} \, \op{\$} \, \op{M}_<^{\otimes 4} \notag \\
&= \frac{1}{32} \bigl( -{\op{Z}\op{Z}\op{Z}\op{Z}} +
{\op{Z}\op{Z}\op{Z}\op{Y}} +{\op{Z}\op{Z}\op{Y}\op{Z}} -
{\op{Z}\op{Z}\op{Y}\op{Y}} \notag \\
& \qquad + {\op{Z}\op{Y}\op{Z}\op{Z}} - {\op{Z}\op{Y}\op{Z}\op{Y}} -
{\op{Z}\op{Y}\op{Y}\op{Z}} +
{\op{Z}\op{Y}\op{Y}\op{Y}} \notag \\
& \qquad + {\op{Y}\op{Z}\op{Z}\op{Z}} - {\op{Y}\op{Z}\op{Z}\op{Y}} -
{\op{Y}\op{Z}\op{Y}\op{Z}} +
{\op{Y}\op{Z}\op{Y}\op{Y}} \notag \\
& \qquad -{\op{Y}\op{Y}\op{Z}\op{Z}} + {\op{Y}\op{Y}\op{Z}\op{Y}} +
{\op{Y}\op{Y}\op{Y}\op{Z}}
-		{\op{Y}\op{Y}\op{Y}\op{Y}} \bigr) \notag \\
& \qquad + \frac{1}{8} \op{I} \op{I} \op{I} \op{I} \notag \\
&=\: -\frac{1}{8} (\op{B}_{zy})^{\otimes 4} + \frac{1}{8} \op{I}^{\otimes 4}.
\end{align}
The first term can be considered to arise as a result of quantum
effects while the identity term is the classical part, since
$\frac{1}{8}$ is just the average payoff in a classical (unentangled)
Minority game when the players select the optimal strategy.
Similarly, $\op{M}_>^{\dagger} \, \op{P}_{0/1} \, \op{M}_> =
\frac{\op{I} \pm \op{B}_{xy}}{2}$, with $\op{B}_{xy} =
\frac{1}{\sqrt{2}} (\op{X} - \op{Y})$.
This expression coincides with
that for the strategy $\op{M}_<$ with the simple substitution $\op{Z}
\leftrightarrow \op{X}$.
Thus for $ \op{\$}_> $ we get expression \eqref{eq:payoff_ob1}
with $\op{Z}$ replaced by $\op{X}$.

Because of the similarity between \eqref{eq:payoff_ob1} and a
four-particle Bell polynomial we write \eqref{eq:payoff_av} as a
\emph{payoff polynomial}
\begin{align}
\label{eq:payoff_poly}
\chi_\text{payoff} &= 4 - E(1111) + E(1112) + E(1121) - E(1122) \notag \\
&\quad + E(1211) - E(1212) - E(1221) + E(1222) \notag \\
&\quad + E(2111) - E(2112) - E(2121)+ E(2122) \notag \\
&\quad - E(2211) + E(2212) + E(2221) - E(2222),
\end{align}
where we have scaled \eqref{eq:payoff_av} by a factor of 32 for
simplicity.  Here $E(k_1, k_2, k_3, k_4) = \bra{\psi_\text{in}} \op{A}_{k_1}
\otimes \ldots \otimes \op{A}_{k_4} \ket{\psi_\text{in}}$
with $k_i \in \{0,1\}$.
In the case of the
QMG, the observables $\{\op{A}_1, \op{A}_2 \} \equiv \{\op{X},\op{Y}\}$ for $\alpha >
\sqrt{\frac{2}{3}}$ and $\{\op{Z},\op{Y}\}$ for $\alpha <
\sqrt{\frac{2}{3}}$, give $\chi_\text{payoff} = 8\alpha^2$
and $\chi_\text{payoff} = 4 - 2 \alpha^2 + 4 \alpha \sqrt{2 - 2 \alpha^2}$,
respectively,
i.e., the payoffs scaled by the factor of 32,
as expected.

Now consider a four-particle MABK Bell polynomial,
which can be written as~\cite{mermin90,ardehali92,belinskii93}
\begin{align}
  \label{eq:mabk}
  \chi_\text{MABK} &= 
-E(1111) - E(1112) - E(1121) + E(1122) \notag \\
&\quad - E(1211) + E(1212)  + E(1221) + E(1222) \notag \\
&\quad - E(2111) + E(2112) + E(2121) + E(2122) \notag \\
&\quad + E(2211) + E(2212) + E(2221) - E(2222).
\end{align}
A local realistic theory must satisfy the Bell inequality
\begin{equation}
\label{eq:mabk_in}
\abs{\chi_\text{MABK}} \le 4.  
\end{equation}
This inequality is maximally violated (by a factor of $\Delta \chi = 2
\sqrt{2}$) by the GHZ state if the first three observers measure in
the $\{ \op{X},\op{Y} \}$ basis and the fourth observer in the
$\{\frac{1}{\sqrt{2}} (\op{X}-\op{Y} ), \frac{1}{\sqrt{2}}
(\op{X}+\op{Y}) \}$ basis.  Alternately, following~\cite{werner01},
maximal violation is obtained if all observers measure in the basis
$\{\cos(-\pi/16) \op{X} + \sin(-\pi/16) \op{Y}, \, \cos(7 \pi/16)
\op{X} + \sin(7 \pi/16) \op{Y} \}$.  These measurement schemes give
maximal violation for the state in \eqref{eq:initial} provided $\alpha
\ge \sqrt{\frac{2}{3}}$.  For $\alpha \le \sqrt{\frac{2}{3}}$, maximal
violation is obtained with the same measurement scheme except with
$\op{X}$ replaced with $\op{Z}$.  The violation $\Delta \chi$ of the
MABK inequality \eqref{eq:mabk_in} versus $\alpha$ is given in
Fig.~\ref{fig:payoffs}.

A comparison of \eqref{eq:payoff_poly} and \eqref{eq:mabk} shows a
striking similarity between the two expressions. They only differ in
the distribution of relative signs ($\pm$) associated with each term
$E(\hdots)$. Remarkably, in the context of the rules and optimal
strategies of the QMG, the payoff polynomial \eqref{eq:payoff_poly}
evaluated with the optimal-strategy measurement schemes $\{
\op{X},\op{Y} \}^{\otimes 4}$ (for $\alpha \ge \sqrt{\frac{2}{3}}$)
and $\{ \op{Z},\op{Y} \}^{\otimes 4}$ (for $\alpha \le
\sqrt{\frac{2}{3}}$) for the initial state \eqref{eq:initial} has the
same value as the MABK polynomial \eqref{eq:mabk} evaluated with these
schemes and this initial state. In a local realistic theory, the
absolute value of both polynomials is always $\le 4$. Since we have
already shown that the strategies \eqref{eq:dflksjg1} are optimal, we
can infer that, for a symmetric strategy profile, the dichotomic
observables $\{ \op{X},\op{Y} \}^{\otimes 4}$ (if $\alpha \ge
\sqrt{\frac{2}{3}}$) and $\{ \op{Z},\op{Y} \}^{\otimes 4}$ (if $\alpha
\le \sqrt{\frac{2}{3}}$) give the maximum violation of the Bell
inequality defined by \eqref{eq:payoff_poly}, i.e.,
$\abs{\chi_\text{payoff}} \le 4$.  Given the continuous spectrum of
initial states \eqref{eq:initial}, optimizing the payoff in the QMG
thus corresponds to maximizing the violation of the Bell inequality
defined by \eqref{eq:payoff_poly}. This observation directly relates
the four-player QMG discussed in this paper to Bell inequalities.


We now show that the Minority game is effectively unique amongst
four-player, symmetric, binary-choice games in its connection to Bell inequalities. 
We restrict ourselves to considering Bell
polynomials with measurement in a single plane (i.e., $\hat{X}\hat{Y},
\hat{X}\hat{Z}, \hat{Y}\hat{Z}$) though there are Bell inequalities
defined on two planes that may also give rise to a connection with
quantum games. We assume that all players choose the same strategy and
consider a general payoff observable
\begin{equation} 
\op{\$} = \sum_{j_1,j_2,j_3,j_4=0,1} c_{j_1j_2j_3j_4}
	\bigotimes_{k=1}^4 \op{P}^{(k)}_{j_k}.
\end{equation}
Application of a strategy $\op{M}(\theta, \beta, -\beta)$
to each qubit transforms the projection operators
$\op{P}^{(k)}_{j_k}$ as
\begin{align}
\op{M}^\dagger \op{P}_{j} \op{M} &=
\frac{1}{2} \op{I} \:+\: \left(\cos^2 \frac{\theta}{2} - \frac{1}{2}
\right) \op{Z} \notag \\
        & \quad + (-1)^{j}\, \frac{1}{2} \sin \theta
                \left( \sin 2 \beta \, \op{X} - \I \cos 2 \beta \, \op{Y} \right).
\end{align}
To correspond to a single-plane Bell polynomial, we eliminate one of
the terms $\op{X}$, $\op{Y}$, $\op{Z}$ with an appropriate choice of
$\theta$, $\beta$.
This can be done by selecting,
\begin{subequations}
\label{eq:conditions}
\begin{align}
  \op{Z} &: \quad \theta = \frac{\pi}{2}, \frac{3\pi}{2}, \quad
  \text{$\beta$ arbitrary}, \\
  \op{X} &: \quad \text{$\theta$ arbitrary}, \quad \beta = 0,
  \frac{\pi}{2}, \pi, \frac{3\pi}{2}, \\
  \op{Y} &: \quad \text{$\theta$ arbitrary}, \quad \beta = 0,
  \frac{\pi}{4}, \frac{3\pi}{4}, \frac{5\pi}{4}.
\end{align}
\end{subequations}
Next, terms in the polynomial containing identity
operators must mutually cancel, save for an ineliminable but trivial
term $\op{I}^{\otimes 4}$.
This gives fourteen constraints on the $c_{j_1j_2j_3j_4}$s,
with the last two conditions in \eqref{eq:conditions}
having in addition $\sum _{j_1j_2j_3j_4} = 0$.
The constraints can be solved in terms of two parameters $a$ and $b$:
\begin{subequations}
\begin{align}
c_{0001} &= c_{0010} = c_{0100} = c_{1000} \notag \\
&= c_{0111} = c_{1011} = c_{1101} = c_{1110} = a \\
c_{0000} &= c_{0011} = c_{0101} = c_{0110} \notag \\
&= c_{1001} = c_{1010} = c_{1100} = c_{1111} = b
\end{align}
\end{subequations}
With $a > 0$ and $b < 0$ we have the Minority game,
while reversing the signs gives the complement,
the \emph{anti-Minority} game,
where each player scores $\frac{1}{n}$ when there is no strict minority,
$n$ being the number of players.
In the case where we use the first of the conditions in \eqref{eq:conditions}
we can have $a$ and $b$ with the same sign,
but these games are the trivial ones where either everyone wins,
or everyone looses,
depending on the sign of $a$.

\section{Extensions}
It is interesting to consider an initial state that, unlike
\eqref{eq:initial}, is completely symmetric with respect to the
interchange of any two qubits,
\begin{multline}
\label{eq:initial2}
\ket{\phi_\text{in}}  = \alpha \ket{\text{GHZ}}
\:+\: \sqrt{\frac{1-\alpha^2}{3}} \left( \ket{\text{EPR}}_{\text{AB}}
  \otimes \ket{\text{EPR}}_{\text{CD}} \right.  \\ 
\left. + \: \ket{\text{EPR}}_{\text{AC}} \otimes
  \ket{\text{EPR}}_{\text{BD}} \:+\: \ket{\text{EPR}}_{\text{AD}}
  \otimes \ket{\text{EPR}}_{\text{BC}} \right).
\end{multline}
The boundary between the EPR- and GHZ-dominated regions now occurs at
$\alpha = \sqrt{\frac{3}{4}}$,
with  expected payoffs $\avg{\$_>} = \frac{1}{4}\alpha^2$
and
\begin{equation}
\avg{\$_<} = \underbrace{\frac{1}{16}}_{\op{I} \op{I} \op{I} \op{I}, \ldots}
	+ \: \frac{3}{16} \Bigl(
		\underbrace{ \frac{2}{3} (\sqrt{3} \alpha \sqrt{1 - \alpha^2}
		+ 1 - \alpha^2)}_{-\op{Y} \op{Z} \op{Y} \op{Z}, \ldots, -\op{Y} \op{Y} \op{Z} \op{Z}, \ldots} \Bigr).
\end{equation}
For this higher-symmetry state, the correspondence between the MABK
inequality and the PO payoffs remains (dashed line in
Fig.~\ref{fig:payoffs}).

The nature of the correspondence for a larger case
is also of interest.
We have considered the particular case of
a six-player QMG using a state analogous to Eq.~\eqref{eq:initial2}.
We find that the switch of optimal strategy occurs at $\alpha =
\sqrt{\frac{6}{19}}$, with payoffs
\begin{subequations}
\begin{align}
\avg{\$_>} &= \frac{2 + 3\alpha^2}{16}; \\
\avg{\$_<} &= \frac{7(2 - \alpha^2)}{64}.
\end{align}
\end{subequations}
Following~\cite{werner01}, the symmetric measurement schemes that give
rise to maximal violation of the MABK inequality for this six-particle
state~\cite{mermin90,ardehali92,belinskii93} are $\{\cos(\pi/24) \,
\op{X} + \sin(\pi/24) \, \op{Y}, \; \cos(13 \pi/24) \, \op{X} +
\sin(13 \pi/24) \, \op{Y} \}$ for $\alpha \ge \sqrt{\frac{5}{13}}$,
and the same with $\op{X} \rightarrow \op{Z}$ for $\alpha \le
\sqrt{\frac{5}{13}}$.  Although for this particular case the switch in
optimal measurement scheme does not occur at the same value of
$\alpha$ as the switch in optimal strategy, the payoff and violation
curves are qualitatively similar. It will be interesting to explore
the trajectory of initial-state entanglement, strategies, and
inequalities which may give rise to direct equivalence for higher-$N$
cases.


\section{\label{sec:conclusion}Conclusions}
We have demonstrated for the first time a direct equivalence between
Bell inequalities and quantum games. The symmetric Pareto optimal payoff
in a four-player quantum Minority game is equivalent to the
violation of the MABK-type Bell inequality for an important class of four-partite
entangled states that involve the GHZ state and products of EPR pairs.
The payoff scheme for the quantum game combined with the strategies
chosen by the players leads to a payoff polynomial that is analogous
to the Bell polynomial.  For both the
optimal payoff in the quantum game and for the Bell inequality there is a {\em quantum fulcrum}
where there is a switch in preferred strategy (quantum game) or
measurement scheme (Bell inequality) corresponding to a change from
the EPR- to GHZ-dominated region.

The equivalence uncovered here is important from the view of both
quantum game theory and Bell inequalities. Our result shows that the
four-player quantum (anti-) Minority game assumes a special position,
since it uses quantum nonlocality to achieve an advantage over the classical case in
precisely the same way nonlocality is evidenced through the
violation of the Bell inequalities.
This also implies that the lessons
learned from interpreting the nature of nonlocality in quantum
mechanics through the lens of Bell inequalities can be readily applied
to advancing our understanding of the discerning features of quantum
game theory over its classical counterpart.

\section*{Acknowledgments}
\noindent
This project was supported by the Australian Research
  Council (ARC). APF and MS are recipients of ARC Postdoctoral
  Fellowships (project numbers DP0559273 and DP0773169, respectively),
  and LCLH is the recipient of an ARC Professorial Fellowship (project
  number DP0770715).

\end{document}